# SEMANTIC WEB SERVICE DISCOVERY APPROACHES: OVERVIEW AND LIMITATIONS

Ibrahim El Bitar, Fatima-Zahra Belouadha and Ounsa Roudies

Department of Computer Engineering, Ecole Mohammadia d'Ingénieurs,
Mohammed V University - Agdal, Rabat, Morocco

## ABSTRACT

*The semantic Web service discovery has been given massive attention within the last few years. With the increasing number of Web services available on the web, looking for a particular service has become very difficult, especially with the evolution of the clients' needs. In this context, various approaches to discover semantic Web services have been proposed. In this paper, we compare these approaches in order to assess their maturity and their adaptation to the current domain requirements. The outcome of this comparison will help us to identify the mechanisms that constitute the strengths of the existing approaches, and thereafter will serve as guideline to determine the basis for a discovery approach more adapted to the current context of Web services.*

## KEYWORDS

*Semantic Web Service Description, Web Service Discovery, Matchmaking, W3C Standards, Case Based Reasoning.*

## 1. INTRODUCTION

Web services (WS) have marked the Web engineering by creating a universal framework that exploits existing Internet protocols and open XML standards to support B2B (Business to Business) interaction [1]. In other words, they are interoperable software components that can be reused in the development of component-based applications and enable their integration.

In this context, assisting the designer or developer in searching for necessary components helps in reducing the cost of developing new applications. Thus, several approaches of WS discovery have been proposed in the literature. Separately, the Web context has evolved with the evolution of the Semantic Web and the increasing adoption of WS as services' implementation technology.

Consequently the number of WS over the Internet is enormously rising. In addition, the evolution of the clients' constraints, those who have become more and more demanding and who constantly try to reuse services so as to meet their needs as well as their functional and non-WS functional requirements.

In this context, our work aims to analyze the proposed approaches in the literature to identify the mechanisms making their strengths, which subsequently will serve as guideline to determine the basis of a new discovery approach that is more adapted to the current WS context.

The remainder of this paper is structured as follow. The second section describes the general context of the WS discovery process. Section 3 highlights the different classes of semantic WS discovery approaches and presents a set of related works. The fourth section provides a comparative evaluation of these approaches based on a set of criteria that can be qualified as performance indicators. The fifth section focuses on the reuse of experience concept that is employed by the existing CBR-based approaches. After that we finely evaluate these specific approaches in section six. The seventh and final section provides a summary of the evaluation and releases a conclusion.

## 2. BACKGROUND AND MOTIVATION

According to the paradigm of WS, their descriptions are published in registries specially designed for this purpose e.g., UDDI (Universal Description Discovery and Integration). These registries aim to facilitate the research of the WS for different commercial organizations wishing to use a particular service. Locating a WS with a particular interest inside the pool of available services is the fundamental task of any WS discovery approach [2]. Otherwise, the WS discovery is the act of locating a machine-treatable description of a WS which is not known before and whose properties meet essentially some precise functional criteria. However, the mechanisms are still needed to ensure effective selection of the appropriate WS instance in terms of quality and performance factors all over the WS consumption [3].

Commonly, WS discovery procedures are based on keyword search and are also guided by manual interference. In response to the requirements identified in his query, the client receives a list of WS descriptions that should be manually scanned to select services that exactly meet its needs. However, within an environment designed for dynamic integration of distributed systems, a fast, automatic and semantic search for composable WS is highly recommended. As shown in Fig. 1, the WS automatic discovery seeks to identify services that can answer a given query by performing a matching of the query elements with the corresponding ones in the descriptions of services stored in a registry. This mechanism usually consists on identifying a similarity degree between semantic concepts describing the required service (query) and the matching ones in the provided services (published services). In the particular case of services described using the WSDL (Web Service Description Language), W3C standard for WS description, the matching covers a set of elements such as operation, input and output.

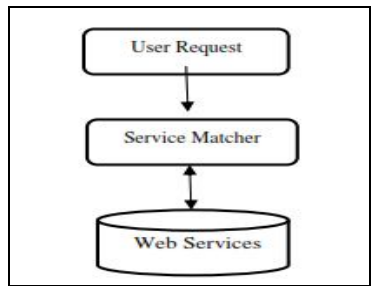

Figure 1. WS Discovery process

Although the field of WS discovery is fairly recent, much work have been devoted to this axis. They have been classified in the literature according to the architecture criterion (centralized, distributed or hybrid), organizational criterion (structural or non-structural) and matching level criterion (syntactic or semantic). However, we find that the proposed classifications are obsolete and focus on secondary endpoints. We believe that these work should be further classified according to the formalism and subsequently the reasoning applied to calculate the matching degree, which is a primary criterion, since the efficiency of the discovery process depends crucially on how the matching is done. This classification of existing work will facilitate the evaluation of their relevance. Our article targets this strategy. We first present a classification of the WS discovery approaches before evaluating their performance by addressing criteria in line with the new requirements of customers and the evolution of Web technologies and standards for semantic WS Description. The focus on semantic WS is not random. It comes obviously, given the current context of the Web developed by the semantic Web and the semantic processing impact on the efficiency of any automatic process.

# 3. SURVEY AND CLASSIFICATION

Our analysis of the WS discovery proposals has enabled us to notice that the adopted approaches have heterogeneities at the matching process level. According to this criterion, they can be grouped into three classes: the algebraic approaches, the deductive approaches and the hybrid approaches. The algebraic approach is an approach that exploits the graph theory and proceeds by calculating distance whereas the deductive approach is essentially based on logic. On the other hand, the hybrid approaches are those that combine the mechanisms of these two classes. Hereinafter, we present the principles of each of these approaches and a set of existing work proposals or platforms that adopt it.

## 3.1. Algebraic Approach

The algebraic discovery approach is based on the calculation of the textual similarity degree from structured graphs built for this purpose, or the calculation of distance (path) between matched concepts. This approach, which we designate as algebraic, uses mechanisms of structural, digital and syntactic matching through a structured graph match and by calculating digital distances to check syntactic similarity. To exploit the semantic, these matching mechanisms use term frequencies and sub-graphs. iMatcher1 [4], AASDU (Agent Approach for Service Discovery and Utilization) [5] and DSD Matchmaker [6] are work proposals and platforms adopting this type of approach.

iMatcher1 [4] is a system of non-logical WS discovery, using service profile syntactic matchmaker. The system picks the services from a set of WS profiles described in OWL-S. These services are stored as RDF (Resource Description Framework) graphs serialized into an RDF database, using an extension of the language RDQL (RDF Data Query Language) called iRDQL [7]. The degree of matching of the query and a service is calculated from four metrics for syntactic similarity calculation: TFIDF (Term Frequency-Inverse Document Frequency) [8], the Levenshtein similarity distance [9][10], the Cosine vector measurement [11][12] and the divergence measurement of Jensen-Shannon [13]. The results are sorted according to the numerical scores of these syntactic similarity measurements and a user-defined threshold.

AASDU is a multi-agent approach proposed for WS discovery [5]. In this approach, the query is expressed as a string via a GUI (Graphical User Interface). It is then sent to a QAA (Query Agent Analyzer). The latter extracts relevant keywords that he will use afterward to select other expert agents from the system repository of the service agents expertise areas. To do this, it performs a syntactic matching based on a simple variant of the technique TFIDF (Term Frequency Inverse Document Frequency) [14]. Subsequently, selected expert agents transmit services' parameters that belong to their field of expertise to the composition agent which invokes a candidate service according to the choice made by the user or compose a new complex service from some of the offered services to respond to the user's request.

DSD matchmaker [6] performs WS discovery through matching service description graphs. These descriptions are specified using the object-oriented language for service description DSD (Document Structure Description) [15], that specifies variables and sets of declarative objects without any logic-based semantics. The matching process determines from a graph the variables to satisfy, and selects based upon the services' status, the one which better responds to the query from the set of discovered services, and returns a numerical value representing the corresponding matching degree. To perform the matching process, DSD matchmaker runs two parallel descriptions (the current offer description and the query description) as trees and recursively compares the nodes of the two graphs [16]. The algorithm [6][17][18] used for this end, calculates the matching degree of two nodes as an aggregation of the matching degree of two types of semantic concepts represented by the two nodes and the degree of matching properties of these two concepts.

## 3.2. Deductive Approach

The deductive approach is based on logics. Work proposals opting for this kind of approaches use services descriptions and queries that are described in languages derived from logical formalisms, such as description logic and first order logic. They also use logical rules to discover WS and exploit ontology in reason to cover their semantic aspect. To calculate the matching degree, they use various methods and focus on different elements of the services descriptions besides taking into account their semantics. They essentially opt for three types of matching: IO-matching (Inputs and Outputs matching) [19][20], PE-matching (Preconditions and Effects matching) [4] and IOPE-matching (Inputs, Outputs, Preconditions and Effects matching) [21][22][23][24]. In IO-matching, matched elements are limited to inputs and outputs. PE-matching chooses preconditions and effects as matching elements. Regarding IOPE-matching, semantic concepts describing the inputs, outputs, preconditions and effects are the purpose of both services and queries matching.

Paolucci and al., proposes a semantic matching of the WSC (Web Services capabilities) [20] based on the use of DAML (DARPA Agent Markup Language) ontology. Published services and queries refer to DAML semantically describing concepts. The implemented match is IO-matching. By matching semantic DAML concepts, Jones considers that a published service will match a request when all query outputs correspond to the published service outputs and all published service inputs correspond to the query inputs. Four matching degrees are offered: EXACT, PLUGIN, SUBSUMES and FAIL. In fact, the outputs are firstly matched and the selected service is the one with the highest score. The input matching is only done in case of an EXACT outputs match. The matching algorithm compares one by one the outputs and the inputs and stores those which most match.

Jaeger and al., offer an OWL-S Matchmaker based on IOPE-matching [21]. The matchmaker researches for semantic mappings between the functional parameters defined in WS OWL-S descriptions and the parameters introduced in the query [25]. The matching process involves four tasks: inputs matching, outputs matching, service categories matching and a fourth task during which constraints and pre-defined features are applied by the user. The algorithm computes the matching degree in each of the first three tasks and performs the fourth one to finally aggregate the results and return back the rank of the matched WS whether it is selected as a service that meets the query. This ranking classes the service according to its matching degree among all discovered WS.

Keller and al., propose a model for automatic and semantic services localization called ALS (Automatic Location of Services) [22]. This model uses goals and predefined abstractions of services to finally discover concrete services responding to a query. The proposed approach distinguishes five levels of matching degrees called intentional matching degrees (Match, parMatch, PossMatch, NoMatch and PossParMatch). GR (Graded Relevances) [23] is an extension of ALS in terms of matching degrees. GR adds two more matching degrees: RelationMatch and ExcessMatch.

Vu and al., have proposed in 2005 a semantic WS discovery approach based on P2P (Pear to Pear) network called PSWSD (P2P-based Semantic Web Service Discovery) [26]. This approach uses a deductive architecture for WS discovery in P2P network. WS Descriptions are provided according to the WSMO ontology using specific techniques and are published in various registries distributed in the P2P network. A user seeking for a WS, with specific QoS (quality of service) constraints can address its query to any registry in the network. The one that received the query forwards it to the registries that can satisfy it. The functionality of the requested service are then extracted from the query and sent to the matchmaker module. The matchmaker selects services descriptions which match semantically with the user query and sends the results to the user who chooses the service to invoke.

CASD (Context Aware Service Discovery) [27] is a further service discovery proposal taking the context into account. The discovery module uses domain ontology to determine the services' categories that have a semantic relationship with the user query. When first the user makes his query (Qusr) in terms of keywords, another richer query (Qctx) is generated and attached to it. Qusr query allows finding WS that have a semantic relationship with the user research terms, while the Qctx query acts as a filter to select the services that match the user's context. Queries made by users are expressed in SPARQL (SPARQL Protocol and RDF Query Language) which is a query language compatible with OWL graphs. The distributed agent architecture for service discovery DCAASD (Distributed Context Aware Service Discovery Agent for architecture) is the distributed version of CASD approach. It was proposed by Schulzrinne and Arabshian [28].

### 3.3. Hybrid Approach

The hybrid approach uses mechanisms deductive methods that integrate distance calculation. The idea is to overcome the limitations of each of these two mechanisms through combination. Several studies [29][30][31][32][33] [34] opt for this approach.

Rey proposes a hybrid approach of semantic WS discovery based on the Description Logics (DL) [29]. The discovery process consists on searching for the best coverage of a description concept using some terminology. For each description Q and terminology T, it looks for a description E, which most covers Q by the use of T [35]. To do this, the author uses the work of Teege [36] who proposes a class of languages for which the difference between concepts can be calculated in the same manner as difference between sets by means of clausal form descriptions. He chooses a representation language from this class to model WS over which he applies this reasoning. To find the best concept coverage by using a terminology, the author uses an algorithm derived from the Hyper-Graph theory. The algorithm performs a search which is equivalent to the search of minimal transversals with minimal cost in a Hyper-Graph where the vertices are the WS and the edges are the query clauses.

OWLS-MX [30] is a hybrid semantic matchmaker which performs IO-matching between OWL-S services' profiles. In addition to the classic matching whose results (Exact or Fail) indicate that paired elements are exactly the same or do not have any relation, the matchmaker offers five other categories of deductive matching (Plug-in, Subsumes and Subsumed-By) and hybrid matching (Logic-based Fail and Nearest-neighbor). They are based on the logic and the use of syntactic similarity. In response to a query, the used matching algorithm returns an ordered set of discovered services to which are associated a matching degree and a syntactic similarity value along with the query. The semantic matching based on the logic of the services' inputs and outputs is particularly supported by the calculation of their syntactic similarity too. In the literature, different variants of the matching algorithm used by OWLS-MX (OWLS-M1, OWLS-M2, OWLS-M3 and OWLS-M4) have been implemented by choosing different syntactic similarity measures.

In addition, OWLS-iMatcher2 [31] is a hybrid discovery approach who opts for both deductive matching of inputs/outputs concepts and algebraic matching used to calculate the textual similarity of names and signatures of services (e.g. SimPack which is intended primarily for the research of similarity between concepts in ontology or ontology as a whole). The aggregation of these two pairings for an overall assessment, uses a binary value having semantic relevance and also different values of similarity metrics proposed in the literature, such as: Bi-gram [38], Levenshtein measurement [39][40] Monge-Elkan measurement [38] and Jaro measurement [38]. The used matching algorithm calculates the syntactic similarity between a given query and all available services then uses a mathematical model regression to predict the matching aggregation value for each service. Discovered services are returned back to the user throughout a decreasing order of the obtained matching values. A statistical evaluation of the returned results as well as their graphical presentation allows the user to interpret and evaluate them easily.

In the FC-Match (Functional Comparison) hybrid approach) [32], the service and query concepts to be matched are presented in OWL-DL (Web Ontology Language Description Logics). The concepts used to describe a service provide information on the elements: Category, Operation, Input and Output. This approach also opts for a deductive semantic matching and an algebraic matching based on the calculation of syntactic similarity. The calculation of the total value matching is made by aggregating the result of the matching based on logical subsumption of concepts between the service and the query and the value of the coefficient of the syntactic similarity.

Meanwhile, WSMO-MX approach [33], is a hybrid approach of matching services described in WSML-MX [41]. She opts for a semantic and syntactic matching and proceeds by using matching techniques for object-oriented graphs inherited from DSD Matchmaker [6] plus the services intentional matching [22]. The overall degree of matching is calculated by aggregation of four values representing the values of the ontological types matching, the logic matching of constraints specified in F-logic (Frame Logic), the matching of the relationship names and the syntactic similarity measurement. The adopted matching centers on the pre-conditions and post-conditions. In addition, the performed semantic matching is categorical. It annotates the obtained results (discovered services) by different values (assumeEquivalent, none, ignore and assumeFailed) according to their degree of matching.

Finally, SAWSDL-MX [34] is a matchmaker inspired from OWLS-MX and WSMO-MX matchmakers which allows discovering the services described in SAWSDL. He opts for both logic matching based on subsumption reasoning and syntactic matching based on information retrieval techniques. The adopted matching covers the following description elements: interface, operation, input and output. To achieve interfaces matching, the matchmaker performs a matching on bipartite graphs (where nodes represent operations and valued arcs whose values are calculated from the matching degree of operations connected by these arcs). In the case of operations matching, it uses various techniques as syntactic approaches (e.g. Loss-of-Information, etc) [42][43], approaches issued from logic (through exploiting semantic annotations of the modelReference attributes contained in the SAWSDL description of the services) and also hybrid approaches. The matching process classes obtained results, according to their degree of matching in the different categories, which are divided into two groups. The first group contains Exact, Plug-in, Subsumes and Subsumed-by; it is used to classify the matching results that are based on logic. While the second group containing Subsumed-by and Nearest-neighbor is used to classify the results of the hybrid matching. Subsumed-by is used when the made matching is not enough and must be supplemented by a syntactic similarity computation. Nearest-neighbor indicates that the false negatives matching were offset in the made matching.

## 4. COMPARATIVE STUDY OF WS DISCOVERY APPROACHES

Using a third approach of hybrid reasoning can deduce that the algebraic and deductive approaches can be combined to meet their limits and consolidate their advantages. By analyzing these three classes of approaches, we can see that the deductive approach applies predefined rules to ensure or not a logical and accurate similarity of a pair of concepts. In the other hand, the algebraic approach calculates the similarity by simply calculating the path (the number of arcs and sometimes the weight of these arcs too) between these concepts (represented as nodes of a graph). We believe that relying on the number of arcs linking a concept to another in a graph is still not a way to accurately determine their degree of similarity and is something that we can reproach for the algebraic approach. However, at another level, we find that for the calculation of the overall similarity between the required and the provided services, this approach uses algebraic formulas to calculate appropriate distances obtained by the aggregation of partial distances. These formulas enable us to obtain measurements whose quality is proven algebraically. Each one of the algebraic and deductive approaches has his own advantage which when combined with the other in a hybrid approach maximizes the accuracy of the calculation of similarity between the required and the provided services.

The classification of existing works proposals in the literature according to the formalism and subsequently the adopted reasoning to perform the matching is certainly an important brick in the process of evaluation. It provides information on efficiency which depends essentially on the matching technique they adopt. However, it is not sufficient to assess their adaptation to the context of WS, therefore other criteria must be taken into consideration.

In this section, we identify five criteria to compare the previously presented approaches. They are mainly related to mechanisms of matching, formalization and description of the WS. These mechanisms are the pillars of any particular approach of WS discovery. The identified comparison criteria include: the matching type, the matching objects, the matching level, the formalism and the service type.

The matching type gives information on its depth. In particular, the concepts used to describe the properties of a WS and those used to express the required properties in a query are not necessarily the same, otherwise they may not be syntactically similar. Also, a deep or "smart" matching should not be limited to the syntactic level. It should cover the semantic aspect or cover both levels if necessary.

The matching objects gives information about the matched elements among those listed in the descriptions of the required WS and the available one. This information is important because it allows us to know on which basis the matching is made. Particularly, the quality of matching results can be evaluated in terms of the elements that have been considered in this matching. A WS whose inputs and outputs correspond to the inputs and outputs of the required WS is not necessarily the appropriate service for the client if it appears that he requires execution preconditions that the first one cannot guarantee. In addition, even if an overall matching of the functional properties set designates a WS as an appropriate service to respond to a query, this cannot be assured especially if the query also includes a set of non-functional constraints required by the client.

The matching level informs about the degree of similarity sought during the matching process. Some work consider that an element matches with its corresponding only in case of high similarity (Exact). Others propose several categories of similarities (Plug-in, subsume, etc.). The matched element does not necessarily have to be the same as its corresponding. It may cover it or be its instance. We believe that this matching flexibility allows, in case of absence of WS that exactly meet a query, to discover others who may partially cover the expressed needs or even encapsulate them.

The formalism criterion informs about the formalism being used to represent WS and queries. In the practice, the way a problem is formalized often determines the type of reasoning to adopt to solve it. The reasoning has a direct impact on the performance but also the quality of results that depend on it. The performance and the deployment of the WS discovery process as well as the relevance of the WS discovered in response to a query depends on the type of reasoning used and recursively the adopted formalism.

The type of services means the ontology or the language used to describe WS. The adopted formalism represents theses services in accordance with the used ontology or language. To achieve semantic matching, most existing work rely on the use of OWL-S or WSMO as semantic models for describing WS. However, it follows that the proposed approach is entirely dependent on the chosen ontology and therefore specific for the discovery of a precise type of WS (as those described in this ontology). These ontologies are research products that have been proposed by research laboratories in order to cover the semantic aspect of the WS. However, WS are in fact published with description files that are conform to W3C standards (WSDL). In order to directly target the services available on the web, we believe that WS representation formalisms must comply with the description standards defined by W3C. In addition, their semantic processing must be independent of any particular type of ontology.

The choice of these criteria is not random. We have identified them based on five metrics that can be considered as main performance indicators of any process, namely, efficiency, effectiveness, flexibility, independence and alignment with standards. In particular, the matching type and the matching object criteria, as they were defined, serve to measure the effectiveness of the discovery process that largely depends on the effectiveness of the matching process. The criterion of formalism is an indicator of its efficiency since the adopted reasoning depends on it. The matching level is a criterion to measure the flexibility of the discovery process. Finally, the type of the target services is an indicator that reveals the work that align with standards. Particularly in the case of WS approaches, the alignment with standards guarantees to these approaches the independence of individual description models (or other ontologies).

Table 1 summarizes the results of a comparative study of the related work that we have presented earlier in this paper. This comparison table allows determining how much each of these works would be qualified as effective, efficient, flexible, autonomous and aligned with standards, through evaluating these features.

Table 1.Comparative evaluation of WS discovery approaches

| | | Formalism | Matching type | Matching objet | Matching level | Service type |
|---|---|---|---|---|---|---|
| **ALGEBRIAC APPROACH** | **iMatcher1** | RDF Graphe | Syntactic | Functional properties (Service Profile) | Exact, Fail | OWL_S |
| | **AASDU** | Set of words | Syntactic | Functional properties (Inputs/Outputs) | Exact, Fail | WSDL |
| | **DSD-Matchmaker** | DSD Graphs | Syntactic | Functional properties (Inputs/Outputs) | Exact, Fail | Diane Service Desciption |
| **DEDUCTIVE APPROA** | **WSC** | DAML Concepts | Semantic | Functional properties (Inputs/Outputs) | Exact, Plug-in, Subsume, Fail | DAML |
| | **OWLS-M** | OWL-S | Semantic | Functional properties (Inputs/Outputs) | Equivalent, Subsume, Unknown, Fail | OWL-S |
| | **ALS** | Logics (Goals + Axioms) | Semantic | Functional properties (Inputs/Outputs/Goals) | Match, PossMatch, ParMatch, PossParMatch, No-M., Excess M. | WSMO |
| | **PSWSD** | WSMO | Semantic | Functional properties (Goals, Preconditions/ Postconditions) | Exact, Fail | WSMO |
| | **CASD** | OWL Graph | Semantic | Functional properties (Inputs/Outputs) | Exact, Fail | CASD |
| **HYBRID APPROACH** | **LD** | Logic Description + Hypergraph | Semantic | Functional properties (Inputs/Outputs) | Exact, Fail | Not specified |
| | **OWLS-MX** | OWL-S | Semantic | Functional properties (Inputs/Outputs) | Equivalent, Subsume, Subsumed by, Logic-Based fail, N.N., Fail | OWL-S |
| | **OWLS-iMatcher2** | OWL-S | Semantic & syntactic | Functional properties (Inputs/Outputs) | Exact | OWL-S |
| | **FC-Match** | OWL-S + OWL-DL | Semantic & syntactic | Functional properties (Inputs/Outputs/ Categories/Operations) | Exact | OWL-S |
| | **WSMO-MX** | WSMO | Semantic & syntactic | Functional properties (Service Goals, Preconditions/ Postconditions) | Equivalent, Plug-in, Inverse Plug-in, Intersection, Fuzzy similarity, Neutral, Disjunction | WSMO |
| | **SAWSDL-MX** | SAWSDL | Semantic & syntactic | Functional properties (Inputs/Outputs/ Operations) | Equivalent, Subsume, Subsumed by, N.N., Fail | SAWSDL |

Moreover, as we can note from the comparison table, the formalism and the reasoning adopted for matching WS are a crucial and primary criterion which is not exclusive to evaluate WS discovery approaches. Other criteria, namely the level, object and type of match as well as the type of target services, also are important to evaluate the adaptation of these approaches in the context of WS. By analyzing these criteria, we can conclude that a suitable WS discovery approach should:

- Target preferably standard WS aligning with W3C standards.
- Integrate non-functional properties to WS functional properties as global matching objects, to better respond to the user requirements.
- Opt for a semantic matching so as to not dispel possible results as in the case of syntactic matching.
- Opt for a matching type that ensures a minimum degree of flexibility. In other words, do not be limited to an exact approach so as to enable the discovery of potential services that can respond to the query even if their properties do not correspond exactly to those required by the query.

Indeed, we have relatively distinguished our work alongside the existing comparative studies [4] [44] through our vision of classification and the criteria we have used to evaluate the WS discovery approaches. We note that these work have not dealt with the "reuse of experience" aspect in the WS discovery. They have excluded an important class of the hybrid approach which consists of the intelligent CBR-based work. We detail this idea in the following section.

# 5. REUSE OF EXPERIENCE AND CBR-BASED APPROACHES FOR WS DISCOVERY

The integration of semantics in the WS description has undoubtedly improved their interpretation and subsequently their discovery process by identifying and selecting the appropriate services. However, integrating semantics does not mean automating the discovery especially with users or admins' intervention to refine results in many existing approaches.

The WS discovery distinctly the WS automatic matching can be improved while taking into account the relevance of research experiences and matching already done for the same requested task, and so giving us valuable information on the behavior of services which is normally difficult to presume before the execution of the service. Therefore, it is necessary to develop a methodology or an approach that uses a knowledge representation of the specific field that concerns the required task in order to capture the WS execution experiences and to use them in the matching process. Punctuality in the selection could and should take into account past successful experiences to reduce the response time and facilitate the matching between retrieved services.

The Case Based Reasoning (CBR) provides such a methodology which is founded on the reuse of experience formed during the resolution of a given problem, in order to resolve a similar one. However, few studies in the literature have focused on optimizing the time of discovery or composition of WS by exploiting the CBR [45] [46] [47] [48] [49] [50] [51]. As follows, we present an overview of theses work in order to identify their characteristics.

Thakker, Osman and Al-Dabass present their S-CBR approach in which the WS execution experiences are modeled as cases that represent the WS properties in a specific area described using OWL semantic description [45][46]. Initially, the administrator performs the repository with semantic formats for case representation in a particular field. This representation is used to semantically annotate the users' queries looking for adequate services as well as the WS execution experiences in the given specific area. The process begins by receiving the description of a client's query. Formerly, S-CBR engine starts looking for suitable services that match with the query. S-CBR system uses frame structures to model its cases. These structures are not generalized and depend heavily on the application domain so that the constituent elements of these structures more precisely "slots" vary from one application to another.

Lajmi and al., have proposed the WeSCo_CBR approach based mainly on ontologies and CBR meant for the WS composition [47][48].They have created an ontology that describes various features of a WS using OWL trying to bring a semi-automatic guidance for the user. Thus, in order to facilitate the processing, they proceed by transforming the user's query into an ontological formula combining a set of ontology concepts defined previously by the authors. For each received new query, the reuse process consists on retrieving similar prior stored cases and eventually evaluating and storing the new case. In WeSCo_CBR, a case comprises the following three elements: a problem, a solution and an evaluation. The discovery of WS meeting the client's needs is accomplished by using similarity measures designed in accordance with the formalization of the problem. The most relevant case is usually determined according to its similarity with the new problem case. In this context, the authors have developed methods not to only calculate similarity between cases but also to guide the research and to determine the most appropriate cases.

In order to improve the WS discovery, Wang & Cao have intervened by introducing an additional CBR-based component called CBR/OWL-S Matching Engine [49]. To better target the desired WS, this matching engine uses ontologies for a semantic research besides a CBR digital evaluation formula (similarity measure). The basis of the CBR/OWL-S matching is the CBR engine which integrates the OWL-S reasoner. After receiving a query, the matching engine selects its suitable cases from the Case Base and then the CBR engine calculates the degree of correspondence.

To improve the lack of precision in WS discovery, Franco De Rosa and De Oliveira have presented their work where they propose to semantically describe the WS and to create databases where to search for cases. These databases are effectively based on the users' profiles [50]. This approach allows users to search for services in a database characterized by a specific SAWSDL semantic annotation integrated by the service designer and the application area expert by means of using OWL ontology references. The system engine searches for services on the base of terms, concepts or research cases. If the service is in line with the client's expectations, it can be stored as a research case in its own profile and thus it can be used again to improve the research in the same field.

## 6. COMPARATIVE STUDY OF CBR-BASED WS DISCOVERY APPROACHES

We have identified five criteria to compare the previously reported CBR-based work. We point out specific criteria to evaluate the approaches based on this type of reasoning. In addition to the criteria of an appropriate WS discovery approach that we concluded in the section 4, we consider specific criteria that are mainly related to the mechanisms of formalization of WS cases and the organization of the case base. These mechanisms are the pillars of any particular WS discovery approach based on CBR. Besides, the comparison criteria that we have identified are: case representation, matching objects, semantic annotation, type of case matching, matching level and finally organization of the WS case base.

The case representation criterion corresponds to the formalization of the WS discovery case. It refers to the data or the knowledge represented in a case. We believe that the case formalization should preferably be aligned with WS standards.

The matching objects criterion reflects the ability to take into account during the formalization of the case, different properties that may constitute the client's query, namely the functional and non-functional properties. The more properties taken into account by the approach, the more it meets the needs of users by considering these properties as matching objects.

The semantic annotation defines the language used to semantically describe the WS and the discovery cases. However, our previous studies have revealed a multitude of languages for WS semantic description and have also noted the advantage of the SAWSDL W3C standard [51] [52]. The question we have to ask is whether the annotations used in these approaches are consistent with this standard or not.

The retrieval of WS cases is the fundamental phase of the CBR-based discovery process. It is based on the case matching which can be syntactic or semantic. The relevance of the results and the discovery process depends on the relevance of the implemented matching algorithm and similarity measures.

The level of case matching informs about the degree of similarity sought in the matching process (Exact, subsumes, etc..).

The cases are classified in a registry called case base. In order to facilitate the search for the most appropriate case to the target problem, it is necessary to organize the case base. The rapidity of the retrieval phase depends on the adopted method of organization. Indeed, the organization of the base must allow faster access to the adequate cases.

Table 2 summarizes the results of a comparative study of the CBR-based work that we have previously presented. This comparative table shows the characteristics of each of these work.

Table 2.Comparison of CBR-based WS discovery approaches

| | Case representation | Matching Object | Semantic annotation | Matching type | Case base organization | Matching level |
|---|---|---|---|---|---|---|
| **S-CBR (2006)** | *(funct prop, non-funct prop, solution)* | - Functional properties (Inputs/Outputs/ constraints/featur e) <br> - Non-functional properties (preferences) | OWL-S | - Semantic (Weighted Block city measure) | Partitioning by characteristics | - Exact <br> - Fail |
| **WeSCo-CBR (2006) (2009)** | *(User profile, Activities, Variables, Instances, set of activities, evaluation)* | Functional properties (Inputs/Outputs) | OWL-S | - Semantic (adapted Manhattan measure) | Partitioning by business area | - Exact <br> - Fail |
| **CBR-OWLS (2007)** | *(funct prop., solution)* | Functional properties (Inputs/Outputs) | OWL-S | - Semantic (Weighted specific measure) | No partitioning | - Exact <br> - Subclass <br> - Superclass <br> - Overlapping <br> - Fail |
| **De Franco Rosa & De Oliviera (2008)** | *(funct prop., solution)* | Functional properties (Inputs/Outputs) | SAWSDL | - Syntactic/ Semantic (Measure not specified) | No partitioning | - Exact <br> - Fail |

The work presented in this section adopt different languages or ontologies to describe the semantic cases. These languages or ontologies vary between OWL-S, SAWSDL or semantic model covering a predefined vocabulary. Each of these work has a case base which must be supplied from a set of source cases reflecting concrete WS. However, the existing registries of WS use practically the WSDL standard. Thus the representation of the cases should preferably be compatible with WSDL. However, only the work presented by De Franco Rosa & De Oliviera [50] uses the SAWSDL standard which allows semantic annotation of WSDL services. We also find that these work do not describe similarly the cases. Each work proceeds by defining its own way to constitute a case relatively to the adopted WS description language, the composition requirements or the target application domain.

In addition, all of these work are interested to the functional properties. In fact, the functional information should be targeted as a priority in the WS discovery. Particularly the S-CBR platform covers non-functional properties in addition to functional properties, in order to better respond to its users exigencies. Subsequently, the performance of any discovery system based on CBR depends on its ability to cover as much WS descriptive properties as possible.

Moreover, as for the matching methods, the studied work adopt different matching techniques based on semantic similarity measures which sometimes integrate the weighting of the matched attributes. This aims to provide a high weight to an attribute compared to another one during the similarity computation. Moreover, the matching set up in the work of [50] consists first of a syntactic search in the case base before performing a semantic search in the UDDI if necessary. The syntactic search does not allow identifying similar cases that do not match syntactically and directs the system to an expensive search in the UDDI.

Finally, the organization of the case base allows a fast selection, and so satisfying the customer in a minimum response time. All of the presented work do not take this aspect into consideration excepting the work of [45] [46] and [47] [48] that offer respectively a partitioning of the case base by features expressed using a predefined vocabulary or by business domain.

In fact, the organization by features enables to search in a partition containing cases having the same feature. However, we should ask whether it is the proper partitioning in the case of WS or not. For example, when the query looks for a service having a given business category and offering a specific functionality, searching in a partition of services that share the same business category is not the most appropriate way. This can be explained by the fact that this partition can contain a multitude of services that offer functionalities other than the requested one even if they belong to the same category.

On the other hand, the organization by business domain partially solves the problem of the selection rapidity, since that the application of such an organization would be beneficial in a case base covering several business domains. However, a hierarchical organization will also be necessary to facilitate the search of the case in a case base relative to one business domain.

# 7. SYNTHESIS AND CONCLUSION

The continuing need to improve the WS discovery involves several approaches. The hybrid approach combines the techniques of the algebraic and the deductive approaches. This approach involves a category of work that adopt artificial intelligence reasoning to guide a dynamic WS discovery by reusing successful experience.

Each work in this category has its advantages and limitations. Definitely, none of them effectively meets all the criteria we have established in our study. Also, any new WS discovery approach, regardless of the vision on which it is based, must benefit from its predecessors' advantages and address their shortcomings in terms of covering the criteria mentioned before.

## Authors

Authors Affiliation: Siweb Research team, Ecole Mohammadia d'Ingénieurs, Mohammed V-Agdal University, Morocco.

### I. EL Bitar

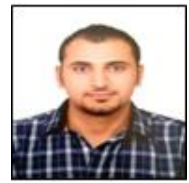

Master Degree in Computer Science and Telecommunications in 2010 with distinction. PhD Student in Computer Science. Lecturer at the Mohammadia School of Engineers (EMI)-Computer Science Departement. Lecturer at the National School of Mineral Industry (ENIM)-Computer Science Departement. 9 recent publications papers between 2011 and 2014; Ongoing research interests: Semantic Web services Discovery, Case Based Reasoning.

### F.Z. Belouadha

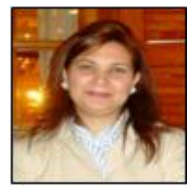

Doctorate degree in Computer Science with distinction in 1999; Former Assistant chief of the Computer Science Department at the Mohammadia School of Engineers (EMI); Associate Professor at the Computer Science Department (EMI); Best paper award at SIIE'08 Conference; Recognition award at IEEE International AICCSA'09 Conference; 15 recent publications papers between 2011 and 2014; Ongoing research interests: Semantic composite Web services, Pervasive information systems/m-services, Business intelligence, MDA .

### O. Roudies

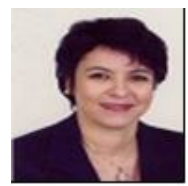

Doctorate degree in Computer Science in 1989, PhD in Computer Science in 2001; Former chief of the Computer Science Department at the Mohammadia School of Engineers (EMI); Chief of Computer Science field at EMI; Professor at the Computer Science Department-EMI; Co-Editor of the eti Journal; 17 recent publications papers between 2011 and 2014; Ongoing research interests: SI, composition, Web services, patterns, Quality.